\documentclass[showpacs,reprint,prb,amsmath,amssymb,floatfix,superscriptaddress]{revtex4-1}

\usepackage{graphicx}
\usepackage{amsmath} 
\usepackage{amssymb}
\usepackage{amsfonts}
\usepackage{bbm}
\usepackage{color}
\usepackage[dvipsnames]{xcolor}
\usepackage{upgreek}

\newcommand{\bone}{\mathbf{1}}
\newcommand{\eqdef}{\overset{\text{def}}{=}}

\begin{document}

\title{Large enhancement of conductivity in Weyl semimetals with tilted cones: Pseudorelativity and linear response}
\author{Saber Rostamzadeh}
\email{srostamzadeh@sabanciuniv.edu}
\affiliation{Faculty of Engineering and Natural Sciences, Sabanci University, Orhanli-Tuzla, Istanbul, Turkey}
\affiliation{Laboratoire de Physique des Solides, Universit\'{e} Paris-Sud, Universit\'e Paris Saclay, CNRS UMR 8502, 91405 Orsay Cedex, France}
\author{\.{I}nan\c{c} Adagideli}
\affiliation{Faculty of Engineering and Natural Sciences, Sabanci University, Orhanli-Tuzla, Istanbul, Turkey}
\author{Mark Oliver Goerbig}
\email{mark-oliver.goerbig@u-psud.fr}
\affiliation{Laboratoire de Physique des Solides, Universit\'{e} Paris-Sud, Universit\'e Paris Saclay, CNRS UMR 8502, 91405 Orsay Cedex, France}

\begin{abstract}
We study the conductivity of two-dimensional graphene-type materials with tilted cones as well as their three-dimensional Weyl counterparts and show that 
a covariant quantum Boltzmann equation is capable of providing an accurate description of these materials' transport properties. The validity of the 
covariant Boltzmann approach is corroborated by calculations within the Kubo formula. We find a strong anisotropy in the conductivities parallel and perpendicular
to the tilt direction upon an increase of the tilt parameter $\eta$, which can be interpreted as the boost parameter of a Lorentz transformation. 
While the ratio between the two conductivities is $\sqrt{1-\eta^2}$ in the two-dimensional case where only the conductivity perpendicular to the tilt 
direction diverges for $\eta\rightarrow 1$, both conductivities diverge in three-dimensional Weyl semimetals, where $\eta=1$ separates a type-I (for $\eta<1$)
from a type-II Weyl semimetal (for $\eta>1$). 

\end{abstract}

\maketitle

\section{Introduction}
Recently, condensed matter systems having Dirac-like linear energy dispersions proved to be a fertile ground to rediscover fundamental particles of nature. The Weyl semimetal (WSM) is a phase of matter which hosts yet another emergent excitation of quantum field theory: a long-sought fundamental particle of nature proposed in 1929 by Herman Weyl as a massless
solution to the Dirac equation.\cite{Weyl1929} Crystals featuring WSM phases, contrary to other types of Dirac matter,\cite{Goerbig2017,jia2016weyl,lv2015experimental} have unique topological features such as: the zero-energy excitations in the semimetallic bulk associated with the chiral Weyl Fermions, having definite handedness near two distinct nodes,\cite{NIELSEN1981219,lv2015experimental,xu2015discovery} whereas 
these nodes are connected only at the boundary of the crystal via peculiar half loop surface states known as the Fermi arcs.\cite{Wan2011prb,Deng2016,Wu2016,cWang2016,jia2016weyl} These materials exhibit intriguing and distinct phenomena when exposed 
to electromagnetic fields, such as chiral anomaly, negative magnetoresistance, and the chiral magnetic effect.\cite{xHuang2015prx,ashby2014chiral,gorbar2014chiral}

Dirac materials, possessing anisotropic and tilted energy cones, where the Fermi surface gains eccentricity and deviates from a standard circular shape, have been reported to exist in the two dimensional organic conductors $\alpha$-(BEDT-TTF$)_2$I$_3$ 
subjected to pressure and uniaxial strain.\cite{Kobayashi2007jpsj,Fukuyama2007jpsj,Goerbig2008prb} In these materials, the Dirac crossing occurs away from the high symmetry points of 
the Brillouin zone and their associated spectrum is modeled with a modified Dirac Hamiltonian. The presence of the tilt largely impacts the magneto-electronic 
and optical properties in these two-dimensional systems with tilted cones.\cite{Kobayashi2007jpsj,Goerbig2008prb,Goerbig2014prb,Goerbig2015prb} Recently, another type of Weyl cone with no relativistic analog, due to the 
violation of Lorentz invariance, in transition-metal 
dichalcogenides hosting Weyl fermions has been reported.\cite{Soluyanov2015,Deng2016,Wu2016,Huang2016,cWang2016,sXu2017,nXu2016,liang2016} As compared to the abovementioned (moderately) tilted cones in so-called type-I WSMs, the cones are now ``overtilted'' such that 
the isoenergy lines are no longer closed ellipses but open hyperbolas. 
The search for this type of WSM, coined type II, is ongoing and some candidate materials have been predicted theoretically.\cite{koepernik2016,autes2016,muechler2016,mccormick2017minimal}

Similar to two-dimensional systems, the tilt of the conical spectrum improves the transport qualities of three-dimensional WSMs by increasing the mobility and conductivity of the carriers. This strongly points out the better electronic and spintronic functionality of the tilted materials compared to other types of Dirac matter. For instance, the reported extremely large and non-saturating magnetoresistance, up to $450,000 \% $ in low fields and 
$13\times 10^6 \%$ in high fields,\cite{ali2014,kumar2017,yWang2016} suggests a suitable magnetic memory and spintronics applications. Other peculiar properties of WSMs with a tilted cone are the large conductivity of about $10^8\;\Omega^{-1}\text{cm}^{-1}$ as well as an enlarged chiral anomaly \cite{kumar2017,lv2017} whose origins remain yet unclear and need 
further theoretical investigation. As evidence to the significance of the Dirac cone tilting, the transport calculations\cite{zyuzin2016,mccormick2017semiclassical,saha2018anomalous,ferreiros2017anomalous} show that the anomalous Hall and thermal Hall conductivity, 
Berry curvature and density of states increase with tilt and peak around the critical tilt value, while optical absorption \cite{mukherjee2017absorption} shows no upper bound. Furthermore, 
the critical angle between the tilt and the magnetic field sets a threshold for the collapse of the Landau level formation and magnetic breakdown.\cite{yu2016predicted,o2016magnetic,arjona2017collapse} 

A very distinct interpretation of tilted Dirac and Weyl systems comes from a relativistic perspective where the tilt is identified as the rapidity of a specific Lorentz transformation.\cite{Goerbig2014prb,Goerbig2008prb} In the presence of external fields exploiting the Lorentz covariance, the tilt parameter becomes an essential variable in identifying a boosted frame where the fields transform trivially and this, in turn, facilitates the study of the problem at hand. \cite{tchoumakov2016magnetic,tchoumakov2017magnetic,Goerbig2008prb} Besides this rich attribute, there is another motivation that appeals to the use of relativistic argumentation in describing the physics of the titled Weyl cone. The direct diagonalization of the Dirac equation in the presence of external fields is a cumbersome task, \cite{lukose2007novel,peres2007algebraic} while by utilizing a relativistic picture and redefining the fields, this problem can be tackled easily.\cite{Goerbig2008prb,Goerbig2014prb,Goerbig2009epl,Goerbig2015prb,o2016magnetic} 
This suggests the use of the covariant formalism that allows for a better understanding of the physics of the tilted Weyl cone phase.

In the present paper, we address the electric transport properties of two- and three-dimensional Dirac systems and WSMs as a function of the tilt parameter. We show
that a relativistic viewpoint in the form of a covariant Boltzmann equation allows for a quantitatively accurate description of the magnetoconductivity in the 
diffusive transport regime. The tilt in the electronic energy dispersion in the general Weyl Hamiltonian, characterized by the tilt velocity $v_0$, is equivalent 
to drift velocity $v_\text{drift}=E/B$ under a suitable Lorentz transformation.\cite{Goerbig2008prb,Goerbig2014prb,Goerbig2015prb} After obtaining 
expressions in the relativistically simplified picture where only the magnetic field is present, an inverse Lorentz boost recovers the results in the original frame. Additionally, we demonstrate and elucidate how the DC 
conductivity increases in moderately tilted type-I WSM in terms of the tilting degree. This large enhancement, which is about 7-10 times larger than in standard 
WSMs without tilt, is the focal point of recent experimental debates concerning electronic transport in WSMs.\cite{shekhar2015extremely,kumar2017,ali2014} 
In order to demonstrate the validity of our approach in terms of the 
covariant Boltzmann equation, we perform conductivity calculations using the Kubo formula. As a result, we find that, apart from quantum corrections at energies close to the 
band-contact points, both approaches show a high quantitative agreement. The quantum corrections happen to be due to the interband coupling which is neglected in the Boltzmann equation; 

The plan of the paper is as follows. In Sec. \ref{sec:02} we discuss the basic construction of the manifestly covariant Boltzmann equation to study transport in the electron's comoving frame of reference which is an inertial frame moving with a velocity equal to the electron's drift velocity relative to the laboratory frame of reference. In Sec. \ref{sec:03} we compute the conductivity of a two-dimensional anisotropic and tilted Dirac cone using the covariant Boltzmann equation as well as Kubo 
formula and compare the two results. In Sec. \ref{sec:04} we repeat the same calculation for a three-dimensional system of type I WSM having tilt in $k_z$-direction and compute the 
longitudinal and perpendicular (to the direction of the tilt) bulk conductivities using covariant Boltzmann formula as well as Kubo formula and then compare them. Furthermore,
we provide a qualitative understanding of our findings for the conductivities by computing 
the tilt-induced renormalization of Fermi velocity, the density of states and Einstein's diffusion relation. 

\section{Covariant Boltzmann equation}
\label{sec:02}

Tilted Dirac cones in a magnetic field, in both two dimensional (2D) and (3D) materials, can be elegantly described within a covariant formulation,\cite{Goerbig2009epl,Goerbig2015prb,tchoumakov2016magnetic}
in which the tilt parameter is associated with an effective electric field.
 
In order to see this, consider first the minimal Dirac Hamiltonian, in which we omit the valley degree of freedom. Let us first consider the manifestly covariant situation with a non-tilted isotropic
Dirac cone in three spatial dimensions. In this case, the covariant Dirac equation can be cast into the Schr\"odinger-type equation, in terms of the Pauli 
matrices $\sigma=(\sigma_x,\sigma_y,\sigma_z)$ and the $2\times 2$ identity matrix $\bone$,
\begin{equation}
 (H-i\frac{\partial}{\partial t})\:|\psi\rangle=0,
\end{equation}
in terms of the Hamiltonian 
\begin{equation}\label{eq:ham1}
 H= v_F (\mathbf{k}-e\:\mathbf{A})\cdot\sigma + e\Phi_{\text{bias}} (\mathbf{r})\bone,
\end{equation}
where $\Phi_{\text{bias}} (\mathbf{r}) = - Ey$ is a static potential that we choose (arbitrarily) to give rise to an electric field in the $y$-direction, while $\mathbf{A}$ is the vector 
potential that generates a homogeneous (and static) magnetic field in the $z$-direction, and $|\psi\rangle=\psi(v_Ft,x,y,z)$. The Fermi velocity $v_F$ plays the role of the speed of light and here, and in the remainder
of the paper, we use a system of units with $\hbar=1$. For convenience, we consider the Landau gauge $\mathbf{A} = - yB\: \hat{\mathbf{x}}$ with 
a vector potential to be oriented in the $x$-direction although the general arguments do not depend on this choice. While the Dirac equation is invariant under Lorentz transformations, it happens that it is convenient, for a quantum-mechanical solution, to use a Lorentz boost (in the $x$-direction) to a frame of reference, where the electric field vanishes, in which 
case we simply need to solve the Hamiltonian for a (charged) Weyl particle in a magnetic field and then transform the solution $\psi'(v_Ft',x',y',z')$ in the comoving frame of reference
back to the laboratory frame, with $\psi'(v_Ft',x',y',z')=S(\Lambda)\psi(v_Ft,x,y,z)$. Notice that this
transformation is only possible in the so-called \textit{magnetic} regime, where the drift velocity $v_D=E/B$ is smaller than the Fermi velocity. It has been shown, both in the 
case of two dimensions \cite{lukose2007novel} and three dimensions \cite{tchoumakov2016magnetic}, that the convenient Lorentz transformations allow for a simple solution of the Hamiltonian (\ref{eq:ham1}),
which agrees with the more cumbersome algebraic solution \cite{peres2007algebraic} in an arbitrary frame of reference, where the electric field does not vanish. 

Interestingly, the same covariant trick can be used for a tilted Dirac or Weyl cone in the presence of a magnetic field. In this case, we need to add the term 
\begin{equation}
 v_0 (k_x -eA_x)\bone, 
\end{equation}
to our Hamiltonian (\ref{eq:ham1}), where we have considered the tilt to be in the $x$-direction, characterized by the velocity $v_0$, which we choose to be smaller than $v_F$ in 
order to remain in the regime of a type-I WSM. One immediately notices that the additional term conspires with the electric field 
in the $y$-direction, and one can indeed define a Hamiltonian 
\begin{equation}\label{Ham_1}
H - v_0 k_x\bone = v_F (\mathbf{k}-e\:\mathbf{A})\cdot\sigma+ e(\Phi_{\text{bias}}-\Phi_{\text{eff}})\bone,
\end{equation}
which has the same form as (\ref{eq:ham1}) in terms of the \textit{effective} electric field $E_0\eqdef v_0 B$ and thus $\Phi_{\text{eff}}=v_0A_x=-E_0y$ . The new Hamiltonian is therefore also of covariant 
form and amenable to a Lorentz transformation, while the extra term $v_0 k_x\bone$ remains diagonal in any frame of reference since $k_x$ is a good quantum number in the Landau
gauge. We, therefore, use, for a convenient diagonalization of the Hamiltonian of tilted Dirac and Weyl cones, a Lorentz transformation to the comoving frame of reference, with
velocity $v_0$ in the $x$-direction, where $S(\Lambda)=\exp(\vartheta \sigma_x/2)$ and rapidity as $\eta=\tanh\vartheta=E_0/v_FB$, to make the effective electric field $E_0$ vanish.\cite{Goerbig2008prb,Goerbig2009epl,Goerbig2014prb,Goerbig2015prb,lukose2007novel}
In the new frame the Dirac equation reads
\begin{equation}\label{eq:Lmap}
(H^\prime-\varepsilon^\prime)\:\;e^{-\vartheta\sigma_x/2}|\psi\rangle=0,
\end{equation}
such that
\begin{align}\label{Ham_2}
H^\prime&=S(\Lambda)\:(H-v_0\:k_x\bone)\:S(\Lambda)\nonumber\\
&=v_F\:(\mathbf{k}^\prime-e\:\mathbf{A}^\prime)\cdot\sigma+e\Phi_{\text{bias}}^\prime(\mathbf{r})\bone,
\end{align}
where $\mathbf{A}^\prime$ and $\mathbf{k}^\prime$ are the corresponding vector potential and 
momentum vector in the comoving frame and $\varepsilon^\prime=\gamma\varepsilon$ (for more details, see the Appendix).\cite{Goerbig2015prb,tchoumakov2016magnetic}
One notices that the spectrum in both the frames are related, i.e.
$\varepsilon(E_0,B)-v_0\:k_x\equiv \sqrt{1-\eta^2}[\varepsilon^\prime(B^\prime,E_0^\prime=0)]$.\cite{Goerbig2015prb,tchoumakov2016magnetic}
This, finally, indicates that the spectrum of a tilted system equivalently can be obtained by first solving the system in a simplified boosted frame and then 
restoring into the original frame by applying an appropriate representation of the inverse Lorentz transformation. 

The tilt term $v_0 \:k_x\:\bone$ in the Hamiltonian (\ref{Ham_1}) renormalizes the scattering time of the carriers, $\tau^{-1}=\sum_\mathbf{k}|V_\text{imp}|^2\:\delta(H-\varepsilon_F)$. The anisotropy of the spectrum redefines the Fermi 
velocity\cite{Goerbig2008prb} (see Sec. \ref{3_c} for more details) such that the ``Fermi's golden'' rule gives $\tau=\tau_0(1+O(\eta^2))$, where $\tau_0$ is the momentum 
relaxation time when the tilt is zero. This recommends that as long as the tilt is moderate, the relaxation time is mainly dominated by impurity scattering.

Furthermore, in the framework of the Liouville equation, the quantum mechanical phase-space distribution function of the carriers of the system (\ref{Ham_1}) satisfies the 
equation $\partial\rho/\partial t+i[H,\rho]=0$ where $\rho$ is the single-particle density matrix.
We next project this equation into the momentum basis $\{|k\rangle\}$ where the Hamiltonian is diagonal and the matrix elements of the density matrix can be written 
as diagonal and offdiagonal parts, i.e, $\rho_{kk^\prime}=\langle k|\rho|k^\prime\rangle=f_k\delta_{kk^\prime}+g_{kk^\prime}$. Consequently, the Liouville equation will be decomposed into two coupled equations in terms of $f_k$ and $g_{kk^\prime}$ which must be solved simultaneously to obtain the quantum kinetic equation written in terms of the diagonal distribution matrix $f_k$.\cite{PhysRevB.96.235134,PhysRevB.96.035106} Next, we convert this equation into an effective semiclassical transport 
equation by taking the Wigner transform of the density matrix $f_k$ and obtain 
\begin{equation}
\partial_tf+i\{H(\mathbf{k}),f\}_{P.B}+i[H(\mathbf{k}),f]=0. 
\end{equation}
Here $f\equiv f(\mathbf{r},\mathbf{k})$ indicates Wigner's (matrix)
distribution function, $\{X,Y\}_{P.B}=\partial_{k_\mu}X\:\partial_{x_\mu}Y-\partial_{x_\mu}X\:\partial_{k_\mu}Y$ 
stands for the Poisson's bracket with $\mu$ and $\nu$ run through the space and momentum dimensions and we adopt Einstein's summation 
convention. \cite{cercignani2002relativistic,de_groot,liboff2003kinetic,vasko2006quantum} The commutator $[H(\mathbf{k}),f]$ is the quantum coherence correction to the 
semiclassical transport equation of the two-band model. In this study, we assume that the perturbations are weak such that the transition between the bands is negligible, 
therefore all the band coherence effects, such as the Berry curvature correction to the carriers band velocity, are ignored and we consider the transport regime where the conduction limits only within a single band.\cite{PhysRevB.77.035110,PhysRevB.96.035106,PhysRevB.84.115209} Note that the additional term to the energy, $v_0 \:k_x\:\bone$, in general changes the 
kinetic equation, however for the spatially homogeneous distribution function which we consider here its effect vanishes as the Poisson bracket $\{H(\mathbf{k}),f\}_{P.B}$ becomes 
itself zero in the absence of external fields. However, by introducing the electromagnetic field through minimal coupling in the Landau gauge, the Poisson bracket will produce two drift terms 
illustrating the Lorentz force driving the motion of electrons and written as, 
$\{H[\mathbf{k}-e\mathbf{A},\Phi(\mathbf{r})],f\}_{P.B}=-e\:(\mathbf{E}+\mathbf{E}_0)\cdot\nabla_\mathbf{k}f- e\:\nabla_\mathbf{k}\varepsilon(k)\times \mathbf{B}\cdot\nabla_\mathbf{k}f$,\cite{PhysRevB.96.235134,PhysRevB.96.035106} where $\nabla_\mathbf{k}\varepsilon(k)$ is the group velocity of the carriers in (\ref{Ham_1}). Therefore, applying the relaxation time approximation, the resulting kinetic equation of the 
tilted system reads
\begin{equation}\label{Boltz_tilt}
\partial_tf+({\mathbf{F}}_\mathbf{E}+{\mathbf{F}}_\mathbf{B})\cdot\nabla_\mathbf{k}f=-\frac{f-f^0}{\tau},
\end{equation}
where $f^0$ is the equilibrium distribution of the carriers and $\mathbf{k}$ is the crystal wave vector of the carrier inside the lattice. For simplicity we separate the Lorentz force into two terms, i.e., $\dot{\mathbf{k}}={\mathbf{F}}_\mathbf{E}+{\mathbf{F}}_\mathbf{B}$, where $\dot{\mathbf{k}}$ represents the time derivative of the crystal momentum, such that
\begin{equation}
\begin{array}{ll}
{\mathbf{F}}_\mathbf{E}=-e\:(\mathbf{E}+\mathbf{E}_0),\\
\\
{\mathbf{F}}_\mathbf{B}=-e\:\nabla_\mathbf{k}\varepsilon(k)\times \mathbf{B}.
\end{array}
\end{equation}
While the second expression is the force caused by the magnetic field, the first drift term is due to the electric fields (bias and effective), and notice that the tilt term (tilt velocity) in the presence of external fields, only enters, through the drift term, into the kinetic part and has no effect on the collision part of the transport equation for the tilted Dirac system. The covariance of the Boltzmann equation indicates that (\ref{Boltz_tilt}) is valid in any frame of reference as
long as one prescribes it in the transformed fields. The equation takes a relatively simple form in a frame of reference where $\mathbf{E}_0=0$.

This, in turn, proves that Hamiltonian (\ref{Ham_1}) and its boosted equivalent (\ref{Ham_2}) yield the same kinetic equation up to the Lorentz symmetry. We then utilize this similarity and compute the physical quantity of interest in an appropriate inertial frame by implementing the coordinate-independent Boltzmann transport equation and then restore to the original frame of reference via the Lorentz transformation law. \cite{cercignani2002relativistic,de_groot,anderson1974relativistic}

The rest of this section is therefore devoted to 
the covariant formulation of Boltzmann's transport equation. The distribution function is a Lorentz scalar since it relates to the number of 
particles, $dN=f(x^\mu,k^\mu)\:dx^\mu\:dk_\mu$, 
through the phase space volume.\cite{juttner1911,cercignani2002relativistic,liboff2003kinetic} Lorentz covariance, as a basic structural 
property of the Dirac equation, can be implemented in investigating the statistical kinetics of the carriers in Dirac (2D) and Weyl (3D) systems. Additionally, throughout our calculations we ignore the spin degree of freedom, as the spin component affected by the Lorentz boost undergoes a precession due to the Wigner rotation\cite{weinberg1995} whereas our main concern is the study of electronic transport of a tilted system. If the electromagnetic fields are present, the ratio between the two fields yields a new velocity that determines the drift of the charged particle, and 
its modulus $v_\text{drift}$ needs to be compared to the effective speed of light $v_F$. If $\eta=v_\text{drift}/v_F<1$, there exists a Lorentz transformation to a frame of 
reference that allows us to get rid of the electric field (\textit{magnetic regime}), while the \textit{electric regime} is associated with a drift velocity that is
larger than the speed of light ($\eta>1$), in which case one can get rid of the magnetic field by an appropriate Lorentz boost. 
In the remainder of this paper, we concentrate on the magnetic regime, which happens to be relevant for the covariant description of type-I Weyl semimetals with moderately
tilted cones.

The covariant Boltzmann equation with electromagnetic fields in manifest covariant form is \cite{liboff2003kinetic,cercignani2002relativistic,anderson1974relativistic}
\begin{equation}
k^\mu \:\partial_\mu f+e\:\mathcal{F}^{\mu\nu}\:k_\nu\:\frac{\partial f}{\partial k^\mu}=-\frac{k^\mu\:u_\mu}{v_F^2}\;\frac{\delta f}{\tau},
\end{equation}
where $\delta f=f-f^0$, is the deviation from the equilibrium distribution. The collision kernel is approximated by a suitable relaxation time ansatz where $u_\mu$ is the four-velocity 
for carrier flow,\cite{anderson1974relativistic,cercignani2002relativistic} which in the electron's local rest frame takes the form $u_\mu=(v_F,0,0,0)$. Using relativistic notations, 
the four-momentum is $k^\mu=(\varepsilon/v_F,\mathbf{k})$ 
where $\varepsilon$ and $v_F$ are the energy and the Fermi velocity of the carriers, respectively. Note that for the massless carriers with Dirac dispersion 
we have $k=\varepsilon v/ v_F^2$, in terms of the four-velocity $v_{\mu}=\dot{x}_\mu=(v_F,\mathbf{v})$, and $\tau$ is the carriers scattering time as the time interval between two successive collisions. Writing the electromagnetic tensor  $\mathcal{F}^{\mu\nu}$ explicitly in terms of the fields $\mathcal{F}^{0i}=-\mathcal{F}^{i0}=E_i/v_F$, 
and $\mathcal{F}^{ij}=-\:\varepsilon_{ij\ell}B_\ell $, the equations of motion read
\begin{align}
k^\mu \:\partial_\mu&=\frac{\varepsilon}{v_F^2}\:(\partial_t+\textbf{v}\cdot\nabla_\mathbf{k}),\\
e\:\mathcal{F}^{\mu\nu}\:k_\nu\frac{\partial}{\partial k^\mu}&=\frac{\varepsilon}{v_F^2}\:e\left(\mathbf{E}\cdot \mathbf{v}\:\frac{\partial}{\partial\varepsilon}+
\mathbf{E}\cdot\nabla_\mathbf{k}+\mathbf{v}\times \mathbf{B}\cdot\nabla_\mathbf{k}\right).
\end{align}
The drift velocity determines the appropriate Lorentz boost defined as (for instance in the $x$-direction of space)
\begin{equation}\label{eq:Lorentz_boost}
\Lambda^\mu_\nu=
\begin{pmatrix}
\gamma&-\gamma\eta& 0 & 0\\
-\gamma\eta& \gamma& 0& 0\\
0&0&1&0\\
0&0&0&1
\end{pmatrix},
\end{equation}
to the frame of reference where $\mathbf{E}_0$ vanishes as long as we remain in the magnetic regime, while the small bias field $\mathbf{E}=\delta\mathbf{E}$ 
is transformed to  $\mathbf{E}^\prime=\delta\mathbf{E}^\prime = \gamma \delta\mathbf{E}_\perp+\delta\mathbf{E}_\parallel$ as well as the magnetic field $B^\prime_z=\gamma^{-1}\:B_z$, ($B^\prime_{x,y}=0$), 
in terms of the Lorentz factor 
\begin{equation}
\gamma^{-1}=\sqrt{1-\eta^2},\qquad \eta=v_\text{drift}/v_F.
\end{equation}
We thus obtain the solution of the covariant Boltzmann equation for a stationary and homogeneous system, the 
nonequilibrium distribution, as\cite{AndoGraphene}
\begin{equation}\label{eq:invariant-dist}
\delta f=\frac{-e\tau}{1+{\omega^\prime}^2\tau^2}\left(\delta \mathbf{E}^\prime+\tau\omega^\prime\:\frac{\mathbf{B}^\prime}{|B^\prime|}\times\delta 
\mathbf{E}^\prime\right)\cdot \nabla_{\mathbf{k}^\prime}\varepsilon^\prime\; \left(\frac{f^{(0)}}{\partial\varepsilon^\prime}\right),
\end{equation}
where, in the comoving frame, the cyclotron frequency of Dirac fermions is given by
\begin{equation}
\omega^\prime=e\:\frac{v_F^2}{\varepsilon^\prime}\:B^\prime.
\end{equation}
The first term inside the parentheses in Eq. (\ref{eq:invariant-dist}) contributes to the longitudinal conductivity whereas the second term gives rise to the Hall current. Throughout this paper we are only interested in the longitudinal conductivity and hence will neglect the effect of the second term in the conductivity calculations.
 
The nonequilibrium current due to an infinitesimal bias is defined through the formula
\begin{align}\label{eq:current}
\mathbf{J}
&=-e\:\text{Tr}(\mathbf{v}\:\delta f),
\end{align}
where the trace (Tr) represents the summation over the momentum and other degrees of freedom. While computing the current in the laboratory frame, we use the Lorentz-invariant
distribution function as calculated in Eq. (\ref{eq:invariant-dist}) prescribed with the transformed fields, namely, 
\begin{align}
\mathbf{\delta E}_\parallel^\prime&= \mathbf{\delta E}_\parallel,\qquad \mathbf{\delta E}_\perp^\prime=\gamma \mathbf{\delta E}_\perp,
\end{align}
and then perform the trace by taking into account that the components of the velocity of the particle in the laboratory frame should transform according to the relativistic 
velocity addition formula, in directions parallel and perpendicular to the boost, as 
\begin{equation}\label{eq:v-addition}
v_\parallel=\frac{v_\parallel^\prime+ v_\text{drift}}{1+\frac{v_\text{drift}}{v_F^2}\:v_\parallel^\prime},\qquad v_\perp=\frac{\sqrt{1-\frac{v_\text{drift}^2}{v_F^2}}}{1+\frac{v_\text{drift}}{v_F^2}\:v_\parallel^\prime}\:v_\perp^\prime.
\end{equation}
Ultimately, one uses the formula $\upsigma_{\mu\nu}=\delta J_\mu/\delta E_\nu$ for the infinitesimal bias fields\cite{CovariantConductivityPRD} to restore the longitudinal conductivity parallel and perpendicular to the boost direction and compare them with the result obtained from the Kubo formula and semiclassical Boltzmann calculations. 

\section{Hamiltonian for 2D anisotropic tilted Dirac cones}
\label{sec:03}

The emergence of Dirac physics in condensed matter makes it possible to probe relativistic effects in materials. In addition, Dirac physics in materials comprises  
electronic properties that are 
generically described by more general Hamiltonians taking into account the anisotropy and the tilt of the energy dispersion. Restricted to two dimensions, the general form of a Dirac Hamiltonian with 
two bands intersecting linearly in a conical shape reads\cite{Goerbig2008prb,Kobayashi2007jpsj}
\begin{equation}\label{eq:hamW}
H_{\text{Dirac}}=v_0^\mu k_\mu\bone+v^\mu k_\mu\sigma_\mu,
\end{equation}
where the indices $\mu$ and $\nu$ run over the spatial dimensions $x,y$.
In addition to the anisotropic Fermi velocity described by the parameter $v^\mu$, 
this Hamiltonian is specified by two parameters, $v_0^x$ and $v_0^y$, that characterize the tilt of the Dirac cones. 
This type of Hamiltonian describes the dispersion of a strained graphene sheet and organic 
conductors of the $\alpha-$(BEDT-TTF$)_2$I$_3$ family.\cite{Kobayashi2007jpsj} 
The additional term $ v_0^\mu k_\mu\bone$ can be understood on the basis of tight-binding models for graphene-like systems. Indeed, second-nearest-neighbor hopping
induces the tilt of the Dirac cones if the latter are dragged away from the $K$ and $K'$ points of the first Brillouin zone (e.g., by strain) and breaks the particle-hole symmetry of 
the system.\cite{goerbig2011electronic,kretinin2013quantum} 

This sufficiently general form of the Hamiltonian (\ref{eq:hamW}) can be simplified by rescaling and rotating in momentum and pseudospin spaces,\cite{Goerbig2014prb,morinari2009possible} to get
\begin{equation}
e^{i\frac{\xi_{\text{tilt}}}{2}\sigma_z}\:H_{\text{Dirac}}(\mathcal{R}^{-1}\mathbf{k})\:e^{-i\frac{\xi_{\text{tilt}}}{2}\sigma_z}\longmapsto  H_{\text{Dirac}}(\mathbf{k}),
\end{equation}
These transformations remove the anisotropy and bring the tilt into the $x$-direction such that one obtains  the minimal Hamiltonian
\begin{equation}\label{weyl}
H_{\text{Dirac}}=v_F\:\mathbf{k}\cdot\sigma+v_0\: k_x \bone.
\end{equation}
where, for simplicity, we set the new Fermi and tilt velocities as $v_F=v_x=v_y$ and $v_0=v_0^x=\eta\: v_F$, respectively. The transformation $\mathcal{R}$ is given by
\begin{equation}
\mathcal{R}(\xi_{\text{tilt}})=
\begin{pmatrix}
\cos\xi_{\text{tilt}}&\frac{v_y}{v_x}\sin\xi_{\text{tilt}}\\
-\sin\xi_{\text{tilt}}&\frac{v_y}{v_x}\cos\xi_{\text{tilt}}
\end{pmatrix},
\end{equation}
where $\xi_{\text{tilt}}=\cos^{-1}\Big(\eta^{-1}v_0^x/v_x\Big)$, and $0<\eta<1$ is the tilt parameter defined as
\begin{equation}
\eta=\sqrt{(v_0^x/v_x)^2+(v_0^y/v_y)^2}.
\end{equation}
Note that the transformation $\mathcal{R}$ in real space is the combination of a coordinate rescaling followed by a pure \text{SO}(2) rotation that maps the original coordinates 
into modified coordinates in order to remove the anisotropy in the $x$-$y$ plane and bring the tilt into a specific direction, 
namely, $(x,y)^\text{t}\longrightarrow (\cos\xi_{\text{tilt}}+i\sigma_y\sin\xi_{\text{tilt}})(x,\frac{v_x}{v_y}y)^t$, and $t$ stands 
for matrix transposition.\cite{Goerbig2014prb,Goerbig2015prb}

Now that we obtained the minimal Hamiltonian model for a general 2D Dirac system, next, we address the transport theory of the system under the influence of electromagnetic fields. Under external fields and in the Landau gauge, as we stated in Sec. \ref{sec:02}, we restore the Hamiltonian (\ref{Ham_1}) for two dimensions. Additionally, in our gauge choice the term $v_0 k_x$ on the left hand side is diagonal in pseudospin and momentum space -- it only shifts 
the spectrum in energy and can thus be absorbed into the Hamiltonian. This indicates that upon minimal coupling the Hamiltonian, $ H_{_\text{Dirac}} - v_0 k_x\bone $, has the same covariant form as that of massless particles in a magnetic field $B\:\hat{\mathbf{z}}$ and an \textit{effective} electric field $E_0$ in the $y$-direction. In the following, we proceed to compute the transport quantities, i.e., longitudinal electric conductivity, by first transforming the system into another inertial frame to remove this effective field.

\subsection{Conductivity from the Boltzmann equation}

The above reasoning allows us to find the Lorentz boost to the appropriate frame of reference, where we can easily calculate the transport coefficients. Notice also that the separation between a strong electric field $E_0$ and a bias field $\delta E$, which was somewhat artificial in the Sec. \ref{sec:02}, is now much more natural -- while $\delta E$ describes a true electric field used to drive a current through the system, $E_0$ simply represents the tilt of the Dirac cones and not a physical electric field. Now assuming that the electric field is oriented in the $y$-direction as in Eq. (\ref{Ham_1}), the bias and magnetic fields under the Lorentz boost in the $x$-direction transform as
\begin{align}\label{eq:bias_transform}
E^\prime_x=\delta E_x,\qquad
E^\prime_y=\gamma\:\delta E_y,\qquad
B^\prime\simeq \gamma^{-1}\:B,
\end{align}
where we have neglected a small correction $\eta \delta E_x/v_F\ll B$ to the magnetic field in the last expression. Additionally, the components of the velocity of the electron in the laboratory frame, using 
Eq. (\ref{eq:v-addition}), relate to those in the comoving frame via 
\begin{equation}\label{eq:v-addition2}
v_x=v_F\;\frac{\cos\phi^\prime+\eta}{1+\eta\cos\phi^\prime},\qquad v_y=v_F\;\frac{\sin\phi^\prime}{\gamma(1+\eta\cos\phi^\prime)}.
\end{equation}
In lights of these relations, due to the relativistic aberration of angles under a boost, the polar angle transforms as
\begin{equation}\label{rel-polar}
d\phi=\frac{d\phi^\prime}{\gamma(1+\eta\cos\phi^\prime)},
\end{equation}
which resembles the relativistic Doppler factor in \textit{relativistic beaming}. This alternatively indicates the Jacobian of the transformation from the lab into the comoving 
frame of reference. Using the transformed form of the velocities $v_\mu$ in Eq. (\ref{eq:v-addition2}), the Lorentz-invariant distribution function (\ref{eq:invariant-dist}) in 
the comoving frame 
and that $\nabla_{\mathbf{k}^\prime}\varepsilon^\prime=v_F(\cos\phi^\prime,\sin\phi^\prime)$, 
one can easily compute the magnetoconductivity tensor in the laboratory frame,
\begin{align}\label{eq:xx_Graphene_conductivity}
J_x&=\frac{-e^2\tau}{2\pi}\int\frac{\varepsilon^\prime d\varepsilon^\prime}{\gamma}\;\left(\frac{\partial f^{(0)}}{\partial\varepsilon^\prime}\right)\nonumber\\
&\qquad\int_0^{2\pi} \frac{d\phi^\prime}{2\pi}\;\frac{(\cos\phi^\prime+\eta)\cos\phi^\prime}{(1+\eta\cos\phi^\prime)^2}\:\delta E_x^\prime,\\\label{eq:yy_Graphene_conductivity}
J_y&=\frac{-e^2\tau}{2\pi}\int\frac{\varepsilon^\prime d\varepsilon^\prime}{\gamma^2}\;\left(\frac{\partial f^{(0)}}{\partial\varepsilon^\prime}\right)\nonumber\\
&\qquad\int_0^{2\pi} \frac{d\phi^\prime}{2\pi}\;\frac{(\sin\phi^\prime)^2}{(1+\eta\cos\phi^\prime)^2}\:\delta E_y^\prime.
\end{align}
Here and in the following parts, we limit our discussion to the diagonal conductivities. Formally, this amounts to considering the limit $\omega'\tau\rightarrow 0$ in Eq. 
(\ref{eq:invariant-dist}), i.e., the limit where the scattering time is much smaller than the inverse cyclotron frequency $\omega^{\prime }$ and where Landau quantization does not
need to be considered. 
In computing the nonequilibrium current in the laboratory frame we, furthermore, need to consider the transformation of the bias fields as given in (\ref{eq:bias_transform}). Now noting the Lorentz transformation of the energy and that at zero temperature the Dirac distribution gives $\left(\partial f^{(0)}/\partial\varepsilon^\prime\right)=-\gamma^{-1}\: \delta(\varepsilon-\varepsilon_F)$, the current and consequently the conductivity perpendicular to the boost direction (which is identified with the tilt direction) in the laboratory frame reads
\begin{align}\label{eq:Boltz_yy}
\upsigma_{\text{perp}}(=\upsigma_{yy})&=\upsigma_0\;\Big(\frac{\gamma-1}{\eta^2}\Big)\:\varepsilon\tau,
\end{align}
where $\upsigma_0=e^2/h$ is a fundamental constant such that $h/e^2=25.8\: \text k \Omega$. Similarly, we find for the conductivity in the direction of the boost (tilt direction)
\begin{equation}\label{eq:Boltz_xx}
 \upsigma_{\text{tilt}}(=\upsigma_{xx})=\upsigma_0\;\Big(\frac{\gamma-1}{\eta^2\gamma}\Big)\:\varepsilon\tau,
\end{equation}
and one thus realizes that the ratio
\begin{equation}\label{eq:ratiosigma}
 \frac{\upsigma_{\text{perp}}}{\upsigma_{\text{tilt}}}=\gamma=(1-\eta^2)^{-1/2},
\end{equation}
is a direct measure of the relativistic factor and thus of the tilt strength $\eta$.

Notice that the above expressions for the conductivity calculated in the framework of the covariant Boltzmann equation coincide, in the limit $\gamma=1$, with those 
obtained previously within semiclassical non-relativistic Boltzmann calculations.\cite{stauber2007electronic} We emphasize that both the relativistic and the non-relativistic Boltzmann 
approaches do not account for quantum corrections in the close vicinity of the band contact points that are responsible, e.g., for the 
minimal conductivity of graphene.\cite{ziegler2007minimal} These are better taken into account within the Kubo formalism, which is discussed in the following subsection. 

\subsection{Kubo formula}

To verify the validity of the covariant approach, we compute the conductivity using the Kubo formula and check the agreement of both methods. Noting that $v_0=\eta\: v_F$, 
the unitary transformation $U=\exp(-i\pi\mathbf{\sigma}\cdot\hat{\mathbf{n}}/4)$, where $\hat{\mathbf{n}}=\hat{\mathbf{z}}\times{\mathbf{k}}/\sqrt{k_x^2+k_y^2}=(\sin\phi,-\cos\phi)$, 
diagonalizes the Hamiltonian (\ref{weyl}), and the spectrum of a general 2D Dirac system thus reads
\begin{equation}\label{polar}
\varepsilon_\alpha=v_F\:k\:(\eta\cos\phi+\alpha),
\end{equation}
where $\phi$ is the momentum vector polar angle and $\alpha=\pm 1$ is the band index. The corresponding Fermi surface is an ellipse. As in the sections above, we consider a 
system with a moderate tilt $\eta<1$, such that we remain in the magnetic regime, where we can get rid of the tilt by the appropriate Lorentz boost. 
In the limit of zero temperature $(T=0)$ for noninteracting systems, the Kubo-Streda formula for the diagonal conductivity gives\cite{akkermans2007mesoscopic}
\begin{equation}\label{kubo}
\upsigma_{\mu\mu}=\frac{e^2}{\pi}\sum_\textbf{k}\int d\varepsilon\:\left(-\frac{\partial f^{(0)}}{\partial\varepsilon}\right)\;\Pi_{\mu\mu},
\end{equation}
where the polarization tensor reads
\begin{equation}
\Pi_{\mu\mu}=\text{Tr}\left(v_\mu\:\text{Im}G^R\:v_\mu\:\text{Im}G^R\right),
\end{equation}
$\text{Im}G^R=(G^R-G^A)/2i$, and the Heisenberg equation gives the velocity operator as $v_\mu=i[\hat{\mathbf{r}}_\mu,H_{\text{Dirac}}]$ where $\hat{\mathbf{r}}_\mu$ is the position operator. In the self-consistent Born approximation, the first contribution to the 
advanced ($A$) and retarded ($R$) Green's function of graphene in the energy eigenbasis reads
\begin{align}\label{green}
G^{^{R/A}}(\varepsilon)=\sum_{\alpha=\pm}\frac{(1+\alpha\:\sigma_z)/2}{\varepsilon-v\:k(\eta\cos\phi+\alpha)\pm i\Gamma},
\end{align}
where the decay term is proportional to the scattering (relaxation) time $\tau$ via $\Gamma^{-1}=2\tau$. Using the above definition, the spectral function is a diagonal matrix in energy basis and may be written as
\begin{align}\label{spectral}
\text{Im}G^R=\frac{1}{\Gamma}
\begin{pmatrix}
\mathfrak{A}_+&0\\
0&\mathfrak{A}_-
\end{pmatrix},
\end{align}
where
\begin{equation}\label{spec}
\mathfrak{A}_\pm=\frac{1}{[z-y(\eta\cos\phi\pm 1)]^2+1},
\end{equation}
in which we defined the dimensionless variables $z=\varepsilon/\Gamma$ and $y=v_Fk/\Gamma$. Now to compute conductivity, we first express the velocity matrix in the helicity basis as
\begin{equation}\label{eq:vlocity_matrix}
v_\mu=ev_F\begin{pmatrix}
v_\mu^{\alpha}&v_\mu^{\alpha\alpha^\prime}\\
[v_\mu^{\alpha\alpha^\prime}]^\ast&v_\mu^{\alpha^\prime}
\end{pmatrix},
\end{equation}
where the band velocities are given by
\begin{equation}
v_\mu^{\alpha}=(\eta+\alpha\cos\phi,\:\sin\phi),
\end{equation}
while the off-diagonal velocities read
\begin{equation}
v_\mu^{\alpha\alpha^\prime}=(i\sin\phi\: e^{-i\phi},\:-i\cos\phi\: e^{-i\phi}).
\end{equation}
We then obtain the polarization as
\begin{align}\label{eq:polarization}
\Pi_{\mu\mu}&=\sum_{\alpha}|\mathfrak{A}_\alpha v^\alpha|^2+\sum_{\alpha\neq\alpha^\prime}\mathfrak{A}_\alpha\:v_\mu^{\alpha\alpha^\prime}\mathfrak{A}_{\alpha^\prime}\:v_\mu^{\alpha^\prime\alpha}\nonumber\\
&=\mathfrak{A}_+^2\:(v_\mu^{+})^2+\mathfrak{A}_-^2\:(v_\mu^{-})^2+2\:\mathfrak{A}_+\mathfrak{A}_-|v_\mu^{+-}|^2,
\end{align}
where the second term in the first lines shows the quantum coherent mixing of the bands, whereas the first term gives the intraband contribution and we remind that $\alpha$ and $\alpha^\prime$ are the band indices. 
Now writing the momentum summation in terms of the momentum and angular integrals as $\sum_\mathbf{k}\rightarrow(2\pi)^{-2}\int_{0}^{\infty}ydy\int_{0}^{2\pi}d\phi$ 
and performing the integrals we finally obtain
\begin{align}\label{eq:Kubo_yy_full}
\upsigma_{yy}&=\frac{\upsigma_0}{\pi}\left[\frac{\gamma-1}{\eta^2}\:(1+z\tan^{-1}z)\right.\nonumber\\
&\quad+ \left.\frac{2z(\tan^{-1}z-z)(1-z/\sqrt{z^2+\eta^2})+\eta^2}{2\eta^2}\right],\\
\label{eq:Kubo_xx_full}
\upsigma_{xx}&=\frac{\upsigma_0}{\pi}\left[\frac{\gamma-1}{\eta^2\:\gamma}\:(1+z\tan^{-1}z)\right.\nonumber\\
&\quad+\left.\frac{\eta^2+2(z-\tan^{-1}z)(z-\sqrt{z^2+\eta^2}}{2\eta^2}\right],
\end{align}
for the diagonal conductivities in the $x$- and $y$-directions. 

Equations (\ref{eq:Kubo_yy_full}) and (\ref{eq:Kubo_xx_full}) are the main result of this section and merit a detailed discussion. First, let us investigate the 
zero-tilt case, $\eta\rightarrow 0$, in which we obtain
\begin{equation}\label{eq:Kubo_notilt}
\upsigma_{xx}=\upsigma_{yy}=\upsigma_0\left[\frac{1}{2\pi}(1+z\tan^{-1}z)+\frac{\tan^{-1}z}{2\pi z}\right].
\end{equation}
In the high-energy (diffusive) limit $z\gg 1$, we then retrieve the standard result, in agreement with the  semiclassical Boltzmann approach, namely\cite{ziegler2007minimal}
\begin{equation}\label{zero-tilt}
\upsigma_{xx}=\upsigma_{yy}=\upsigma_0\:\varepsilon\tau/2,
\end{equation}
which coincides with Eqs. (\ref{eq:Boltz_yy}) and (\ref{eq:Boltz_xx}) in the static limit ($\omega=0$). 
We emphasize that in the opposite limit of $z\ll1$, i.e., at energies close to the Dirac point, band mixing yields 
quantum corrections that are beyond the reach of the semiclassical Boltzmann approach, within the first Born and relaxation time approximation, and therefore a full quantum treatment is needed. 
Indeed, the Kubo and Boltzmann approaches yield different results then, as we will show below. 

\begin{figure}
\centering
\includegraphics[width=1\linewidth]{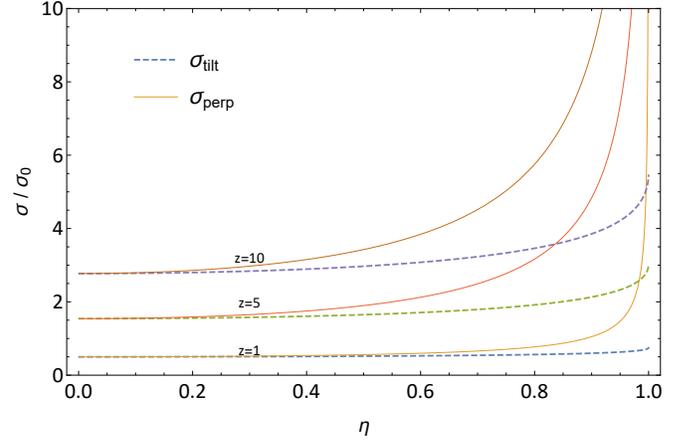}
\caption{(Color online) Comparison between the conductivity of the tilted graphene in directions perpendicular and parallel to the tilt  as a function 
of the tilt parameter $\eta$ for values of the normalized energy as $z= 1, 5$, and $10$. Around zero tilt both the parallel and perpendicular conductivities, 
calculated from the Kubo formula, correspond to the isotropic value; while increasing the tilt develops an anisotropy such that the perpendicular conductivity diverges, while  
the parallel one continues to grow up to a finite value.}
\label{fig:graph-tilt-perp(kubo)}
\end{figure}

\begin{figure}[!]
	\centering
	\includegraphics[width=1\linewidth]{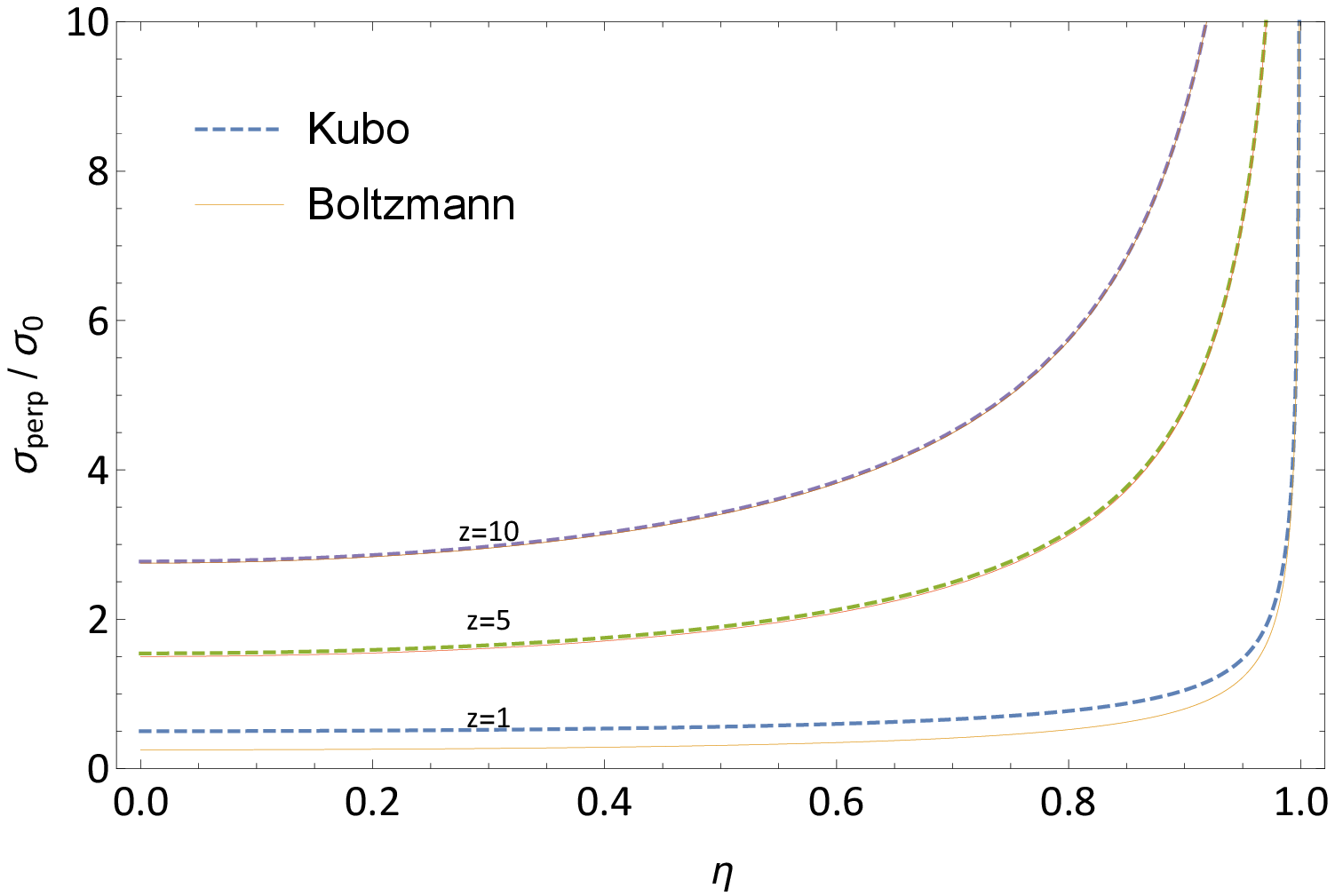}
	\includegraphics[width=1\linewidth]{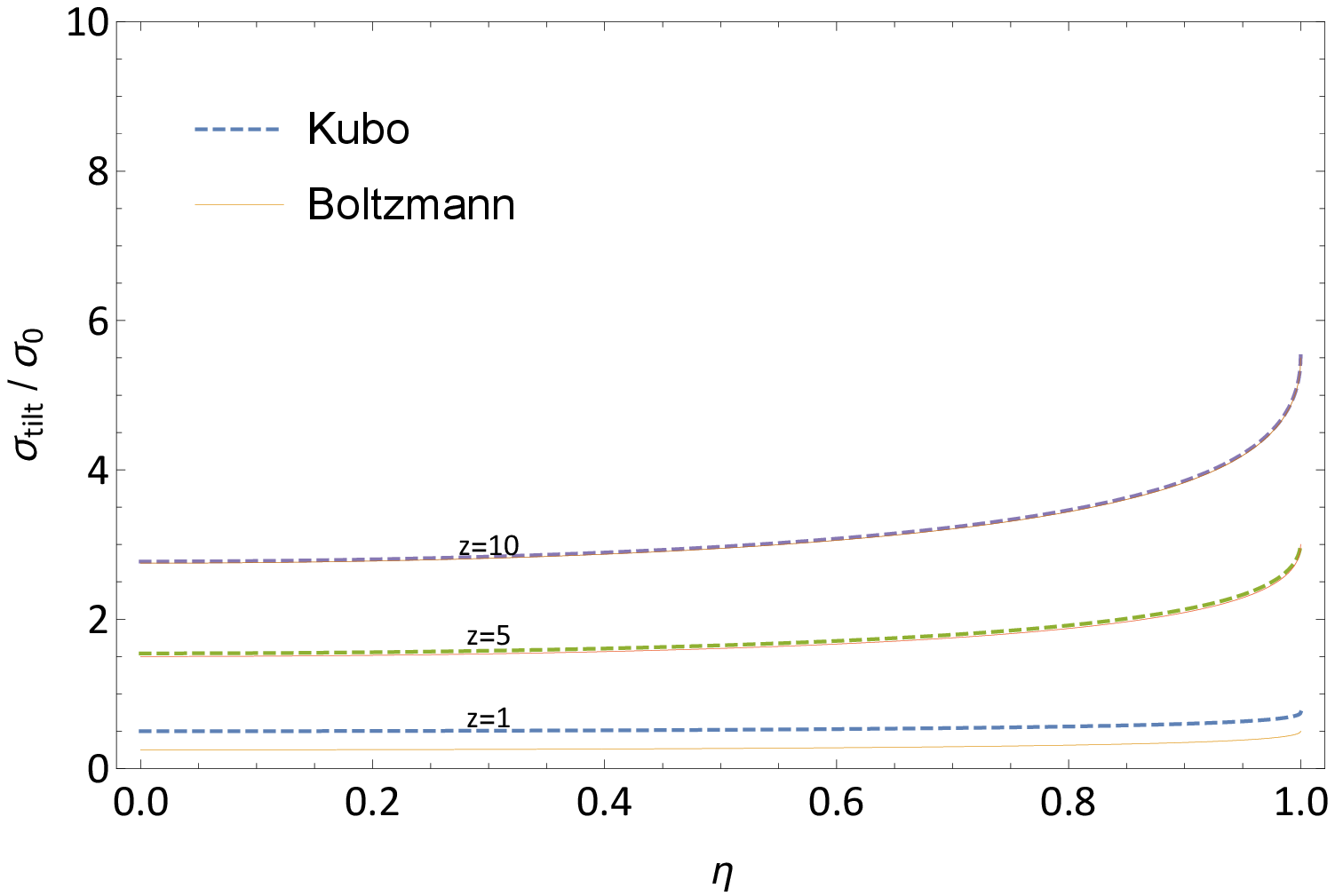}
	\caption{(Color online) Comparison between the conductivities $\upsigma_{\text{tilt}}/\upsigma_0$  and $\upsigma_{\text{perp}}/\upsigma_0$ of the tilted graphene in terms of the tilt parameter for energies at $z=1, 5$, and $10$. Conductivity calculated from the Kubo formula (dashed) agree with the covariant Boltzmann approach (solid) in (\ref{eq:Boltz_xx}), and (\ref{eq:Boltz_yy}). While the perpendicular conductivity diverges in critical limit, the conductivity parallel to tilt direction saturates at finite value.}
	\label{fig:graph-kubo-boltz}
\end{figure}

Before comparing both approaches, let us discuss in detail some aspects of the conductivities calculated from the Kubo formula.
For the case of non-zero tilt in the diffusive limit, $z=\varepsilon/\Gamma=2 \varepsilon\tau \gg 1$,
the conductivities (\ref{eq:Kubo_yy_full}) and (\ref{eq:Kubo_xx_full}) can be rewritten as
\begin{align}\label{eq:Kubo_xx_diff}
\upsigma_{\text{tilt}}(=\upsigma_{xx})&=\upsigma_0\;\Big(\frac{\gamma-1}{\eta^2\:\gamma}\:\frac{z}{2}+\frac{1}{4 z}\Big),\\
\label{eq:Kubo_yy_diff}
\upsigma_{\text{perp}}(=\upsigma_{yy})&=\upsigma_0\;\Big(\frac{\gamma-1}{\eta^2}\:\frac{z}{2}+\frac{1}{4 z}\Big),
\end{align}
where the last term takes into account the first (quantum) correction in $1/z$.  The conductivities (\ref{eq:Kubo_xx_diff}) and (\ref{eq:Kubo_yy_diff}) 
are plotted in Fig. \ref{fig:graph-tilt-perp(kubo)} as
a function of the tilt parameter $\eta$, for different values of $z$. First, notice that the result coincides with that in Eq. (\ref{zero-tilt}) for zero tilt, i.e., 
in the limit $\eta\rightarrow 0$, where we retrieve also $\upsigma_{xx}=\upsigma_{yy}$. Figure \ref{fig:graph-kubo-boltz} shows a comparison between the conductivities 
(\ref{eq:Kubo_xx_diff}) and (\ref{eq:Kubo_yy_diff})
obtained from the Kubo approach (dashed lines) and those from the covariant Boltzmann formula, Eqs. (\ref{eq:Boltz_xx}) and (\ref{eq:Boltz_yy}). The only difference resides 
in the offset of $\delta\upsigma =\upsigma_0/4z=\upsigma_0/8 \varepsilon\tau$, which is due to the quantum corrections that are neglected in the latter approach. 
This offset becomes less relevant at larger values of the tilt where the conductivities are enhanced. This enhancement can globally be understood from the density of states that
increases with the tilt parameter. However, this argument in terms of the density of states does not explain the strong anisotropy in the conductivities. While the conductivity
$\upsigma_{\text{tilt}}$ along the tilt direction remains finite and saturates at a value of 
\begin{equation}
 \upsigma_{\text{tilt}}(\eta\rightarrow 1)=\upsigma_0\left(\varepsilon\tau +\frac{1}{8\varepsilon\tau}\right),
\end{equation}
i.e., the Boltzmann contribution is doubled with respect to the $\eta\rightarrow 0$ limit (\ref{zero-tilt}),
the conductivity in the perpendicular direction diverges\cite{suzumura2013effect}  as 
\begin{equation}
 \upsigma_{\text{perp}}(\eta\rightarrow 1)\sim \upsigma_0 \varepsilon\tau /\sqrt{1-\eta^2}.
\end{equation}
One thus obtains the same behavior (\ref{eq:ratiosigma}) for the ratio between the conductivities as in the Boltzmann analysis.

This difference in the conductivities
stems from the anisotropy of the velocities and mobilities in two directions. Experimentally observed results show that when applying strain to the graphene 
crystal, the group velocity of the carriers along the strain drops by increasing the strain whereas the group velocity along the perpendicular direction to the strain 
increases.\cite{choi2010effects}  The tilt affects the carrier density by increasing the number of available states,\cite{Goerbig2008prb} but ultimately in the surge of scattering events 
in a tilted (elliptic) isoenergy surface, carrier scattering into a state along the perpendicular direction (ellipse shorter axis) is much more probable than along the tilt 
(the larger axes) due to the smaller momentum difference. Hence, the mobility of the carriers is largely influenced by the  direction of the tilt.\cite{cheng2017anisotropic} 
This result is inline with another empirical finding showing that the magnetoresistance takes its maximum measure along the tilt direction in $\alpha$-(BEDT-TTF$)_2$I$_3$ 
organic conductors.\cite{morinari2009possible}

\section{Type-I Weyl Semimetal}
\label{sec:04}

\subsection{Boltzmann equation}

In this section, we present the core result of the paper and investigate the magnetoconductivity of type-I Weyl semimetals based on the covariant formalism introduced in  
Sec. \ref{sec:02}. Consider a Weyl material in the presence of perpendicular electric and magnetic fields -- again, we choose the magnetic field to be oriented in the $z$-direction
and a tilt in the $x$-direction, i.e., the associated electric field is oriented in the $y$-direction. The essential part of the 3D Hamiltonian of this system, according to (\ref{Ham_1}), can be cast into the form $\tilde{H}_{\text{Weyl}}- v_0 k_x\bone$, and,  similar to the discussion in Sec. \ref{sec:02}, has a Lorentz symmetric equivalent.

We apply a Lorentz boost in the $x$-direction with drift velocity $v_0=E_0/B$ to work in a frame where the electric field vanishes. Noting the 
representation of the Lorentz boost, the bias fields transform along the boost direction as $\delta E_{y(z)}^\prime=\gamma\delta E_{y(z)}$, $\delta E_{x}^\prime=\delta E_{x}$ 
and the magnetic field as $B_z^\prime=\gamma^{-1}\:B_z$. Taking the axis of the polar coordinate, $\theta=0$, along the boost direction, simplifies the expressions and using this parametrization in the comoving frame we obtain
$\nabla_{\mathbf{k}^\prime}\varepsilon^\prime=v_F(\cos\theta^\prime,\:\sin\theta^\prime\cos\phi^\prime,\:\sin\theta^\prime\sin\phi^\prime)$. 
Next, implementing the relativistic addition formula, the velocities transform accordingly as
\begin{align}
v_x&=v_F\:\frac{\eta+\cos\theta^\prime}{1+\eta\cos\theta^\prime},\\
v_y&=v_F\: \frac{\sin\theta^\prime\cos\phi^\prime}{\gamma(1+\eta\cos\theta^\prime)},\\
v_z&=v_F\: \frac{\sin\theta^\prime\sin\phi^\prime}{\gamma(1+\eta\cos\theta^\prime)}.
\end{align}
Considering the relativistic aberration of the polar angle $\theta$ under the Lorentz boost, (\ref{rel-polar}), then the solid angle and consequently, the conic cross section transforms as
$d\Omega=\gamma^{-2}d\Omega^\prime/(1+\eta\cos\theta^\prime)^2$ where $d\Omega^\prime=d(\cos\theta^\prime)d\phi^\prime/(4\pi)$ and the azimuthal angle as $d\phi=d\phi^\prime$. Thus,
the nonequilibrium current in the laboratory frame for each spatial direction will give
\begin{align}\label{eq:xx-current_Weyl}
J_x&=\frac{-e^2\tau}{2\pi^2}\int\frac{{\varepsilon^\prime}^2 d\varepsilon^\prime}{v_F}\;\left(\frac{\partial f^{(0)}}{\partial\varepsilon^\prime}\right)\nonumber\\
&\qquad\int \frac{d\Omega^\prime}{\gamma^2}\;\frac{(\cos\theta^\prime+\eta)\cos\theta^\prime}{(1+\eta\cos\theta^\prime)^3}\:\delta E_x^\prime,\\\label{eq:yy-current_Weyl}
J_y&=\frac{-e^2\tau}{2\pi^2}\int\frac{{\varepsilon^\prime}^2 d\varepsilon^\prime}{v_F}\;\left(\frac{\partial f^{(0)}}{\partial\varepsilon^\prime}\right)\nonumber\\
&\qquad\int \frac{d\Omega^\prime}{\gamma^3}\;\frac{\sin^2\theta^\prime\cos^2\phi^\prime}{(1+\eta\cos\theta^\prime)^3}\:\delta E_y^\prime,\\\label{eq:zz-current_Weyl}
J_z&=\frac{-e^2\tau}{2\pi^2}\int\frac{{\varepsilon^\prime}^2 d\varepsilon^\prime}{v_F}\;\left(\frac{\partial f^{(0)}}{\partial\varepsilon^\prime}\right)\nonumber\\
&\qquad\int \frac{d\Omega^\prime}{\gamma^3}\;\frac{\sin^2\theta^\prime\sin^2\phi^\prime}{(1+\eta\cos\theta^\prime)^3}\:\delta E_z^\prime.
\end{align}
As a result, we obtain in the laboratory frame for the conductivity parallel to the boost 
\begin{align}\label{eq:Bplz weyl tilt conduct}
\upsigma_{\text{tilt}}&=\upsigma_0\:\frac{{\varepsilon}^2\tau}{\pi\:v_F}\;\left(\frac{\tanh^{-1}\eta-\eta}{\eta^3}\right).
\end{align}
Similarly for the conductivity in the direction perpendicular to the boost we get
\begin{align}\label{eq:Bolz wyl perp conduc}
\upsigma_{\text{perp}}&=\upsigma_0\:\frac{{\varepsilon}^2\tau}{\pi\:v_F}\;\left(\frac{\gamma^2\eta-\tanh^{-1}\eta}{2\eta^3}\right).
\end{align}
In the zero-tilt limit, $\gamma\rightarrow 1$, we obtain the isotropic conductivity as $\upsigma_{\text{tilt}}=\upsigma_{\text{perp}}=\upsigma_0\:\varepsilon^2\tau/(3\pi v_F)$, recovering the same result as obtained using the semiclassical Boltzmann equation in zero temperature reported elsewhere in the literature. \cite{kim2014boltzmann,burkov2011topological,lundgren2014thermoelectric}

\subsection{Kubo formalism}

\begin{figure}[h!]
	\centering
	\includegraphics[width=0.7\linewidth]{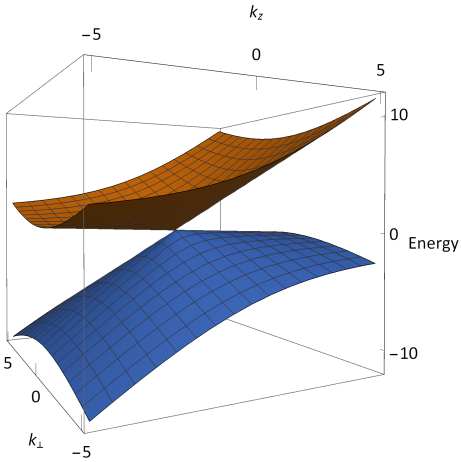}
	\caption{The conical spectrum of a tilted Weyl crossing.}
	\label{fig:tiltedWeylspec}
\end{figure}

\begin{figure}[htp!]
	\centering
	\includegraphics[width=0.9\linewidth]{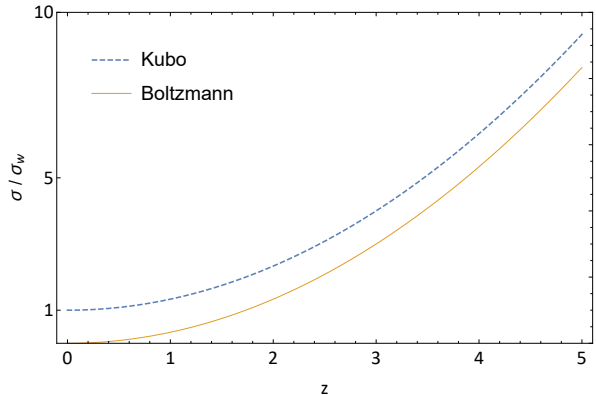}
	\caption{(Color online) Isotropic conductivity  $\upsigma_w=\upsigma_0\frac{\Gamma}{4\pi\: v_F}$ of Weyl semimetals with zero tilt ($\eta=0$) as a function of 
	the normalized energy $z$. The results calculated from the Boltzmann and Kubo approaches agree except for a small offset. This nonzero residual conductivity is due to the band coherence contributions.}
	\label{fig:04}
\end{figure}

Again, we compare the result obtained from the covariant Boltzmann equation to the conductivity calculated within linear response theory. We write a minimal type-I Weyl 
semimetal Hamiltonian with a tilt in the $z$-direction (Fig.\ref{fig:tiltedWeylspec}), parametrized by $\eta=v_0/v< 1$ as
\begin{equation}\label{type-i wsm}
H=v_F \;\mathbf{k}\cdot\sigma+\eta\:v_F\:k_z.
\end{equation}
Similar to the case of graphene, the unitary transformation $U=e^{-i\frac{\theta}{2}\sigma\cdot\hat{\mathbf{n}}}$, where $\hat{\mathbf{n}}=\hat{\mathbf{z}}\times\mathbf{k}/\sqrt{k_x^2+k_y^2}$, brings the Weyl Hamiltonian into the diagonal form. In this
basis, the Green's function through the polar parametrization reads
\begin{align}
G^{^{R/A}}(\varepsilon)=\sum_{\alpha=\pm}\frac{(1+\alpha\:\sigma_z)/2}{\varepsilon-v\:k(\eta\cos\theta+\alpha)\pm i\Gamma},
\end{align}
and the spectral function, in terms of the dimensionless variables, give a similar diagonal form as defined in Eq. (\ref{spectral}) with
\begin{equation}
\mathfrak{A}_\pm=\frac{1}{[z-y(\eta\cos\theta\pm 1)]^2+1}.
\end{equation}
Writing the velocity operators in the energy basis, as given in Eq. (\ref{eq:vlocity_matrix}),
we obtain the band diagonal velocities (using spherical coordinates)
\begin{equation}\label{band-v}
v_\mu^{\alpha}=(\alpha\sin\theta\cos\phi,\:\alpha\sin\theta\sin\phi,\:\eta+\alpha\cos\theta),
\end{equation}
and the off-diagonal velocities are
\begin{align}\label{off_vx}
v_x^{\alpha\alpha^\prime}&=e^{-i\phi}(\cos\theta\cos\phi+i\sin\phi),\\
v_y^{\alpha\alpha^\prime}&=e^{-i\phi}(\cos\theta\sin\phi-i\cos\phi),\\
\label{off_vz}
v_z^{\alpha\alpha^\prime}&=-e^{-i\phi}\sin\theta.
\end{align}
To compute the conductivity using Eq. (\ref{kubo}), we write accordingly the conductivity as
\begin{equation}
\upsigma_{\mu\mu}=\frac{e^2\:\Gamma}{\pi\:v_F}\;\langle\Pi_{\mu\mu}\rangle,
\end{equation}
where the polarization tensor is defined as before in Eq. (\ref{eq:polarization}) with the velocities given in Eqs. (\ref{band-v})-(\ref{off_vz}). Moreover, we
define the three-dimensional momentum and angular integrals now  as $\langle\cdots\rangle=(2\pi)^{-3}\int_{0}^{\infty}y^2 dy\int_{0}^{\pi}\sin\theta d\theta\int_{0}^{2\pi}d\phi$. One
thus finds the conductivity 
\begin{align}\label{eq:kubo wyl conduc}
\upsigma_{\text{perp}}(=\upsigma_{xx}&=\upsigma_{yy})=\upsigma_0\;\frac{\Gamma}{2\pi\:v_F}\big(a_1\:z^2+a_2\big),\\
\upsigma_{\text{tilt}}(=\upsigma_{zz})&=\upsigma_0\;\frac{\Gamma}{2\pi\:v_F}\big(a_3\:z^2+a_4\big),
\end{align}
where the second term yields the (intrinsic) quantum correction to the conductivity due to the band mixing effects,\cite{hosur2012charge} 
and the first term is the Drude conductivity which coincides with the result of the Boltzmann equation in the diffusive limit $z\gg 1$. The other parameters are defined as
\begin{align}
a_1&=\frac{\eta-(1-\eta^2)\tanh^{-1}\eta}{2\eta^3(1-\eta^2)},\\
a_2&=\frac{\eta+(1-\eta^2)\tanh^{-1}\eta}{2\eta(1-\eta^2)},\\
a_3&=\frac{\tanh^{-1}\eta-\eta}{\eta^3},\\
a_4&=\frac{\tanh^{-1}\eta}{\eta},
\end{align}
and we have restored $\upsigma_0$ by noting that $h=2\pi$. In the limit of zero tilt ($\eta\rightarrow 0$), we obtain the isotropic conductivity
\begin{equation}\label{eq:kubo-zerotilt}
\upsigma_{xx}(=\upsigma_{yy}=\upsigma_{zz})=\upsigma_0\:\frac{\Gamma}{4\pi\:v_F}(z^2/3+1),
\end{equation}
recovering the results obtained in previous works using linear response calculations.\cite{ashby2014chiral,gorbar2014chiral,tabert2016optical} In Fig. \ref{fig:04}, we compare the conductivity of a type-I WSM for zero tilt with Fermi velocity $v\approx c/300$, where $c$ is the speed of light, and constant relaxation time $\tau=10^{-7}$s using the Boltzmann approach and Kubo formalism.  
As in the 2D case, we notice a significant 
increase in the conductivities with the tilt in the case of moderate tilt. The discrepancy between the Kubo and Boltzmann conductivities is inherited from the zero-tilt case, 
where we have already noticed that the Kubo formula systematically yields a larger conductivity. In the same manner as in the 2D case discussed above,
the main sources of this discrepancy are quantum interference effects between the two bands 
that are correctly taken into account via the Kubo formula whereas they are not treated yet in the first order linear approximation of the Boltzmann equation and thus are not present in its results. In the zero-tilt limit, the (isotropic) conductivity shows a parabolic behavior in energy that  
can be understood qualitatively from the behavior of the density of states for Weyl semimetal, which is
quadratic in energy. 

\begin{figure}[htp!]
	\centering
	\includegraphics[width=1\linewidth]{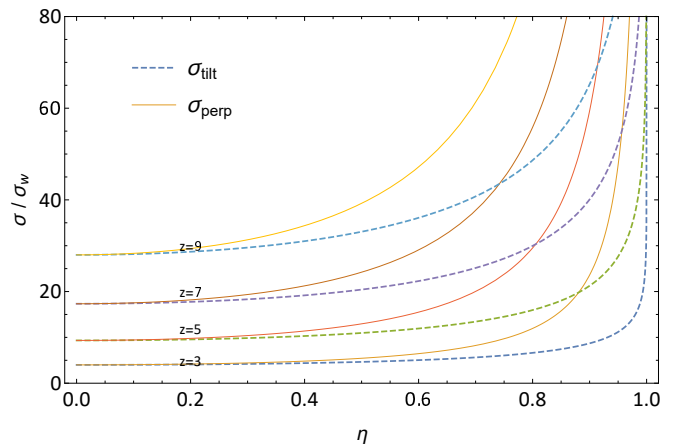}
	\caption{(Color online) Comparison between the parallel-perpendicular conductivity of type-I WSM computed from the Kubo formula in terms of the tilt parameter, 
	for different values of the normalized energies. The conductivity in both directions diverge for $\eta\rightarrow 1$, but the divergence is more pronounced in the direction
	perpendicular to the tilt. In the zero-tilt limit, both conductivities match restoring the isotropic value.}
	\label{fig:06}
\end{figure}

In the nonzero tilt limit, we  find again, from our calculations based on the Kubo formula, that the conductivity $\upsigma_\text{perp}$ 
in the direction perpendicular to the tilt increases and diverges for $\eta\rightarrow 1$ and that the conductivity $\upsigma_\text{tilt}$ in the tilt direction 
is smaller than $\upsigma_\text{perp}$. However, contrary to the 2D case, $\upsigma_\text{tilt}$ now also diverges upon increasing tilt (Fig.\ref{fig:06}). 
The parallel conductivity can be linked to the longitudinal magnetoresistivity if we set the magnetic field in the $z$-direction. Therefore the finite value of conductivity in 
the direction parallel to the tilt is a theoretical evidence of the enhanced longitudinal magnetoresistance observed in WSM having tilted Weyl points.\cite{shekhar2015extremely}
 
This anisotropy in the conductivity can be understood from the covariant point of view where the tilt velocity is identified with an effective electric field normal to 
its direction, $\mathbf{v}_0=\mathbf{E}_0\times\mathbf{B}/B^2$. Thus in effect, the auxiliary field enhances the conductivity along the field (perpendicular to the tilt direction). 

To further corroborate our approach in terms of the covariant Boltzmann equation, we further inspect the anisotropy and the directional features in the conductivity 
in the framework of a semiclassical \textit{noncovariant} Boltzmann equation. In this case, we take into account the tilt-induced anisotropy directly in the expression for the averaged velocities, without appealing to Lorentz boosts to a frame of reference, where the dispersion becomes effectively isotropic. We can then
write the conductivity as
\begin{equation}\label{semicl_Boltz}
\upsigma_{\mu\nu}=\frac{e^2\tau}{2\pi^2 v_F^3}\:\left\langle \frac{v_\mu v_\nu}{(1+\eta\cos\theta)^3} \right\rangle_{\Omega},
\end{equation}
in terms of the band velocities given in Eq. (\ref{band-v}) where $\langle\cdots\rangle_{\Omega}$ denotes the averaging over the random solid angles. 
In performing the angular integrals, the velocity $v_z$ compensates partially for the diverging behavior of the density of states and, as a result, yields a less divergent 
expression for $\upsigma_{zz}$. In contrast, the $v_x$ and $v_y$ velocities 
produce a more singular result and thus yield a strongly divergent behavior
of $\upsigma_{xx}$ and $\upsigma_{yy}$. The calculated conductivity using the non-covariant Boltzmann equation then gives the expressions
\begin{align}\label{eq:Weyl_semicl_conduct}
\upsigma_{xx}=\upsigma_{yy}&=\upsigma_0\:\frac{\varepsilon^2\tau}{\pi v_F}\:\frac{\eta-(1-\eta^2)\tanh^{-1}\eta}{2\eta^3(1-\eta^2)},\\
\upsigma_{zz}&=\upsigma_0\:\frac{\varepsilon^2\tau}{\pi v_F}\:\frac{\tanh^{-1}\eta-\eta}{\eta^3},
\end{align}
which agree with (\ref{eq:Bolz wyl perp conduc}) and (\ref{eq:Bplz weyl tilt conduct}) obtained from the covariant Boltzmann approach. This
further confirms the accuracy of the results obtained from the Kubo formula and covariant Boltzmann approach. 

\begin{figure}[htp!]
	\centering
	\includegraphics[width=1\linewidth]{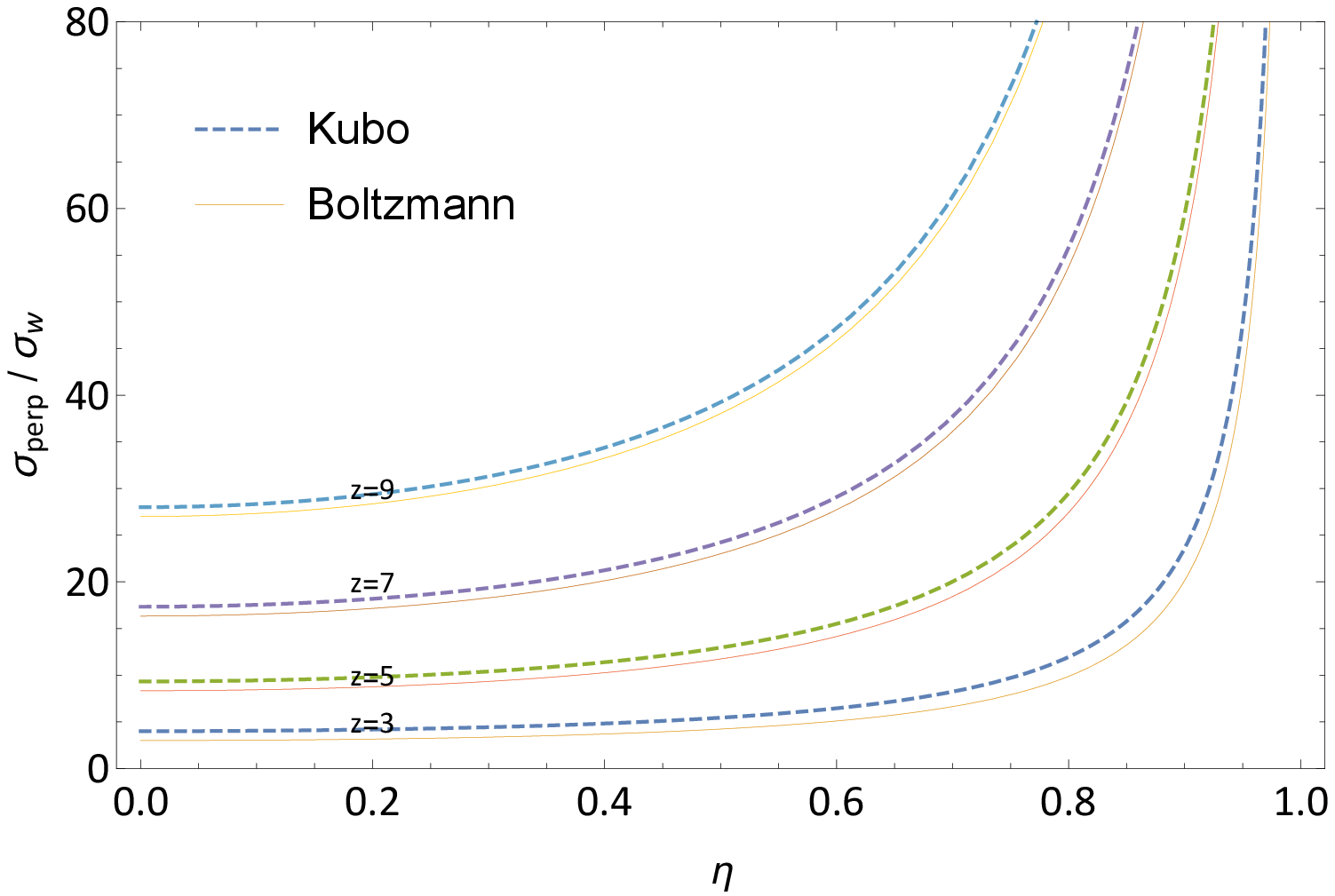}
	\includegraphics[width=1\linewidth]{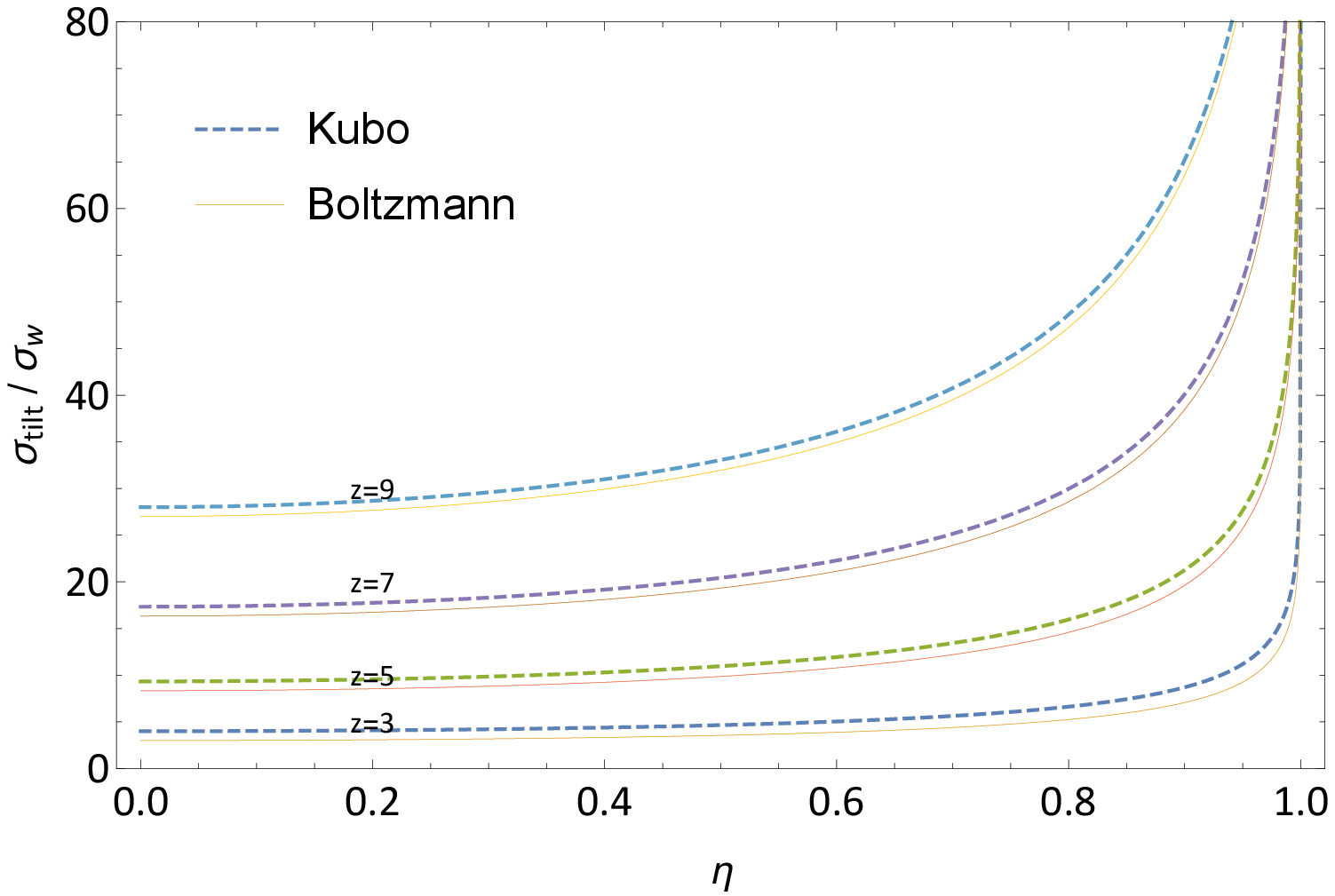}
	\caption{(Color online) Conductivity of the tilted type-I WSM perpendicular and parallel to the tilt direction computed from the covariant Boltzmann equation (solid) and 
	Kubo formula (dashed). The conductivities are expressed as a function of the tilt parameter for different values of the normalized energy $z$. 
	The perpendicular conductivity enhances and diverges at critical value $\eta=1$ while the parallel increases but stays finite in the critical limit.}
	\label{fig:07}
\end{figure}

\begin{figure}[htp!]
	\centering
	    \includegraphics[width=1\linewidth]{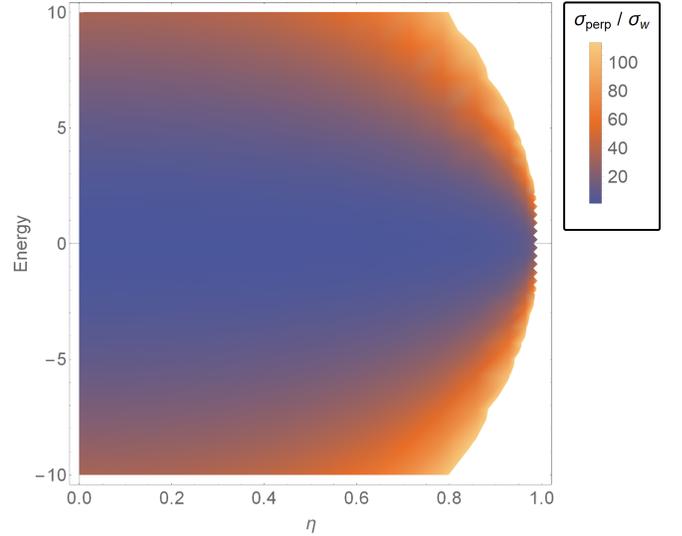}
		\caption{(Color online) Density plot of the normalized perpendicular conductivity in terms of the tilt degree and chemical energy. }
		\label{fig:08}
\end{figure}

As in the 2D case and the above-mentioned zero-tilt limit for Weyl semimetals, we, therefore, find a high  degree of accuracy between the Kubo-formula approach and the results 
from the covariant Boltzmann equation over the whole range of tilt parameters $\eta$. 
This is summarized in Fig. \ref{fig:07}. Again, the main difference stems from the conductivity offset due to interband contributions that are neglected in the Boltzmann
approach. Overall, we find that the increase in the conductivities with the tilt parameter is more pronounced at larger energies, i.e., upon increasing $z$. 
This is also represented in the form of a color plot in Fig. \ref{fig:08}, where we plot the conductivity $\upsigma_\text{perp}$ in the plane spanned by the tilt parameter and 
the energy.

\subsection{Density of states as a function of the tilt parameter}\label{3_c}
\begin{figure}[b]
	\centering
    \includegraphics[width=1\linewidth]{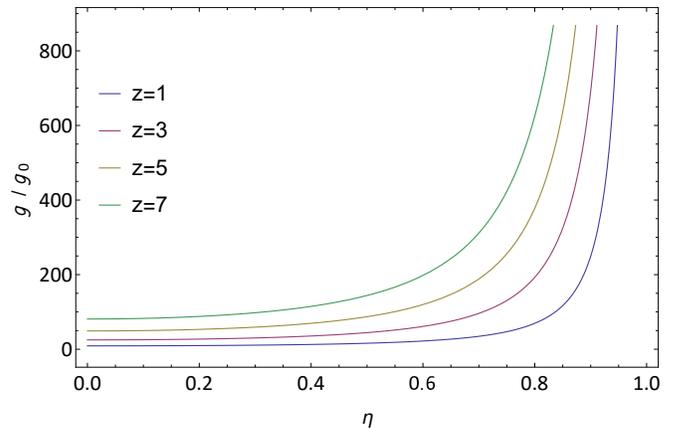}
	\caption{Increasing pattern of the normalized Density of states $g(z)/g_0$ of type-I WSM by the degree of tilting, $g_0=\frac{\Gamma^2}{2\pi^2\:v_F^3}$.}\label{fig:09}
\end{figure}

\begin{figure}[h]
	\centering
	\includegraphics[width=1\linewidth]{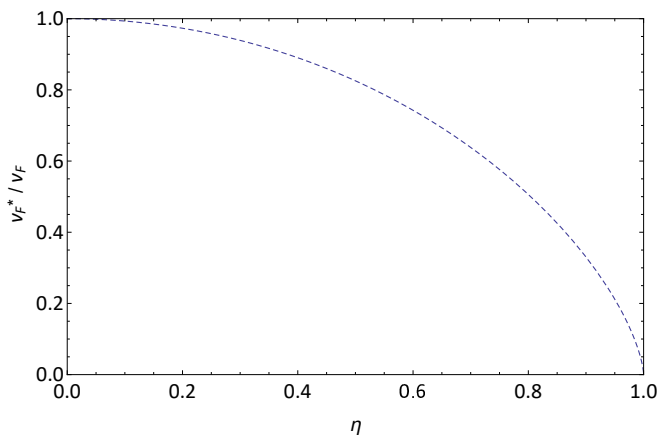}
	\caption{Renormalization of the Fermi velocity of type-I WSM with the tilted cone is a clear indication of the Lorentz violation.}\label{fig:10}
\end{figure}

The increase of the conductivity with the tilt of the Weyl cones can be understood qualitatively from an analysis of the density of states (DOS), $g(\varepsilon)$, which enters into the expression of 
the conductivity in the Einstein relation for a system of $d$-dimension,
\begin{equation}
 \upsigma_{E} = \frac{e^2\bar{v}^2 \tau}{d} g(\varepsilon)\propto g(\varepsilon), 
\end{equation}
that we use here for qualitative analysis. The parameter $\bar{v}$ represents an average velocity since the Einstein relation does not make a difference between 
the directions contrary to the more appropriate Boltzmann or Kubo formulas. 
The tilt of the Weyl cones enlarges the Fermi surface and thus increases the DOS. This yields eventually an enhanced  
conductivity in the type-I WSM.  The hike of the DOS can be justified using both Lorentz covariance and Sommerfeld expansion. From the covariance point of view and 
due to the length contraction in laboratory frame in the direction of boost, it is easy to see that the particle density scales as $dn=\gamma \:dn^\prime$. Furthermore, since $n=\int d\varepsilon\:g(\varepsilon)\:f(T\approx0)$, and noting the transformation of energy, the DOS in laboratory frame, by applying the inverse Lorentz boost, then gives $g(\varepsilon)=\gamma^2\:g^\prime(\varepsilon^\prime)$.
Therefore the expression for the DOS of Weyl semimetals in the laboratory frame reads 
\begin{equation}\label{eq:DOS}
g(\varepsilon,\eta)=\frac{\gamma^4}{2\pi^2}\:\frac{\varepsilon^2}{v_F^3},
\end{equation}
which is plotted in Fig. \ref{fig:09}. One notices that the behavior of the  DOS reflects indeed, as  expected, that of the conductivities. However, we insist that our 
argument in terms of the DOS is qualitative and not sufficient to explain the anisotropy in the conductivities $\upsigma_\text{tilt}$ and $\upsigma_\text{perp}$. 
Notice that the increase of the DOS as a function of the tilt parameter can be absorbed into a renormalized Fermi velocity (see Fig. \ref{fig:10}), with
\begin{equation}
\frac{v_{F}^\ast}{v_F}=(1-\eta^2)^{2/3}.
\end{equation}

In general, the DOS can be calculated by counting the number of occupied states below an energy level. For the WSM having the dispersion (\ref{type-i wsm}) we obtain
\begin{align}\label{dos1}
g(\varepsilon)&=\sum_{\mathbf{k}}^{|k|<k_F}\delta\big(v_Fk(\eta\cos\theta+\alpha)-\varepsilon_F\big)\nonumber\\
&=\frac{\varepsilon^2}{4\pi^2\:v_F^3}\int_{0}^{\pi}\frac{\sin\theta d\theta}{(\eta\cos\theta+\alpha)^3}.
\end{align}
In the limit of $0<\eta<1$ this integral yields the same result (\ref{eq:DOS}) as that obtained from the covariance point of view.

\section{Conclusion}
\label{sec:05}

In conclusion, we have studied the influence of the tilt in the dispersion of graphene (or two-dimensional graphene-like systems) and type-I WSM on the magneto-conductivity. 
The tilt can be described elegantly 
within a covariant framework of the Dirac equation since it can be viewed as an effective electric field that conspires with the magnetic field. In the case of type-I 
WSM with moderate
tilts, $\eta<1$, a Lorentz boost into the comoving frame of reference characterized precisely by the tilt velocity $v_0$ allows one to get rid of the electric field -- this is
the so-called magnetic regime in electrodynamics. This relativistic effect, accompanied by a Lorentz transformation back to the laboratory frame, yields an increased conductivity
that we have studied here within the covariant form of the Boltzmann transport equation both in two and three spatial dimensions. We have systematically compared the results of 
the conductivity from the covariant Boltzmann equation to calculations within the Kubo formula of linear response theory. Furthermore, the conductivities perpendicular 
to the tilt direction, calculated within the Boltzmann and Kubo approach, are enhanced by an extra factor, both in  two- and in  three-dimensional cases, as compared to 
the direction parallel 
to the tilt. This demonstrates that transport in a tilted system becomes directional where $\upsigma_{\text{tilt}}<\upsigma_{\text{perp}}$. This smaller conductivity along 
the field direction ($z$-direction) agrees with the experimental evidence on the extremely large and nonsaturating longitudinal ($z$-direction) magnetoresistance reported 
on type-I WSM. 

Our findings can qualitatively be understood with the help of the DOS, to which the conductivities are roughly proportional, within the simplified picture provided by 
Einstein's relation. The tilt enhances the DOS by some power of the relativistic Lorentz factor $\gamma=1/\sqrt{1-\eta^2}$, which generally enters the expressions, for the 
conductivities on the one hand and for the DOS, scattering time and effective Fermi velocity on the other hand. We further observe that the power of the 
Lorentz factor that enters in the  expressions of the (surface) conductivities, depends on the orientation of the tilt with respect to that of the conductivity as well as on the 
system's spatial dimension. For the renormalization of the effective Fermi velocity, for example, we find $v_{F}^\ast=v_F\:\gamma^{-\beta}$ with $\beta_{2\text{D}}=3/4$ 
and  $\beta_{3\text{D}}=2/3$. Thus measuring the ratio $v_{F}^\ast/v_F$ experimentally would allow for an experimental determination of the tilt and its magnitude in a type-I WSM.

\begin{acknowledgments}
We are grateful to F. Pi\'{e}chon for many insightful and fruitful discussions. We acknowledge financial support from ANR ``SOCRATE'' under Grant No. ANR-15-CE30-0009-01. 
\end{acknowledgments}

\appendix*
\section{Lorentz transformation of the Dirac Hamiltonian}
In this appendix, we provide details for the Lorentz transformation (LT) of the Hamiltonian (\ref{Ham_1}). As mentioned in the main text, we redefine this Hamiltonian as $\tilde{H}=H-v_0 \:k_x\:\bone$ and, for convenience, set $\Phi=\Phi_{\text{bias}}-\Phi_{\text{eff}}$. Using (\ref{eq:Lmap}) and (\ref{Ham_2}), the eigenvalue problem after applying a pure Lorentz boost in $x$-direction, where $\eta=\tanh\vartheta$, gives
\begin{align}\label{eq:eigen}
&\Big(e^{\vartheta\sigma_x/2}\:\tilde{H}\:e^{\vartheta\sigma_x/2}-\varepsilon\:e^{\vartheta\sigma_x}\Big)\;|\tilde{\psi}\rangle=\nonumber\\
&\Bigg\{\!\gamma\!\begin{pmatrix}
e\Phi+v_F\eta\pi_x-\varepsilon\!\!\!&\quad v_F\pi_x-\eta\varepsilon+e\eta\Phi\\ \\ v_F\pi_x-\eta\varepsilon+e\eta\Phi\!\!\!&\quad e\Phi+v_F\eta\pi_x-\varepsilon
\end{pmatrix}\!\!+\!v_Fk_y\sigma_y\!\Bigg\}\!|\tilde{\psi}\rangle,\nonumber\\
&=\Big(v_F(k^\prime_x-eA^\prime_x)\sigma_x+v_F\:k^\prime_y\sigma_y+e\:\Phi^\prime-\varepsilon^\prime\Big)|\tilde{\psi}\rangle,\nonumber
\end{align}
where the kinetic momentum is $\boldsymbol{\pi}=\mathbf{k}-e\mathbf{A}$ and $|\tilde{\psi}\rangle=e^{-\vartheta\sigma_x/2}|\psi\rangle$. In the last line, we have used the LTs
\begin{equation}
k_x^\prime=\gamma(k_x-\eta\frac{\varepsilon}{v_F}),\quad 
\varepsilon^\prime=\gamma(\varepsilon-v_0k_x)=\gamma\varepsilon,
\end{equation}
and note that $v_0=\eta v_F$. Moreover, the four-potential $A^\mu=(\Phi/v_F,\mathbf{A})$ likewise transforms as
\begin{equation}
\begin{pmatrix}
A^{\prime\:0}\\A^{\prime\:1}\\A^{\prime\:2}\\A^{\prime\:3}
\end{pmatrix}=
\begin{pmatrix}
\gamma&-\gamma\eta&0&0\\
-\gamma\eta&\gamma&0&0\\
0&0&1&0\\
0&0&0&1
\end{pmatrix}
\begin{pmatrix}
A^{0}\\A^{1}\\A^{2}\\A^{3}
\end{pmatrix},
\end{equation}
where $\gamma=(1-\eta^2)^{-1/2}$. We thus obtain
\begin{align}
\Phi^\prime&=\gamma (\Phi-\eta v_F A_x),\\
A_x^{\prime}&=\gamma (A_x-\eta \frac{\Phi}{v_F}),
\end{align}
and $A_{y(z)}^{\prime}=A_{y(z)}=0$. Furthermore, noting that 
$\mathbf{v}_\text{boost}=\mathbf{v}_\text{drift}=\mathbf{E}_0\times\mathbf{B}/B^2$, for $\mathbf{E}_0=v_0B\:\hat{\mathbf{y}}$ and $\mathbf{B}=B\:\hat{\mathbf{z}}$, 
one finds that $\mathbf{v}_\text{boost}=\mathbf{v}_\text{drift}=v_0\:\hat{\mathbf{x}}$.
The rapidity of the transformation is also defined as $\eta=\tanh\vartheta= E_0/v_F\:B$. In magnetic regime $\eta\delta E/v_F\ll\eta E_0/v_F<B$, and in the Landau gauge, we ultimately obtain
\begin{align}
\Phi_{\text{eff}}^\prime&=\gamma(\Phi_{\text{eff}}-\eta v_F\: A_x)=\gamma (-E_0+v_0B)y=0,\\
\Phi_{\text{bias}}^\prime&=\gamma\:\Phi_{\text{bias}}=-\gamma\: \delta E y,\\
A_x^\prime&=\gamma(A_x-\eta \Phi/v_F)\approx-\gamma(B-\eta E_0/v_F)y,\nonumber\\
&=-\gamma^{-1}By.
\end{align}
This implies that the boost removes the effective field in the Hamiltonian (\ref{Ham_1})
such that now in the boosted frame we simply have
\begin{align}
\tilde{H}^\prime= v_F\:(\mathbf{k}^\prime-e\:\mathbf{A}^\prime)\cdot\sigma+e\Phi_{\text{bias}}^\prime.
\end{align}

\bibliographystyle{apsrev4-1}
\bibliography{Referances}

\end{document}